\documentclass{article}
\usepackage{graphicx}
\usepackage{amssymb}
\usepackage{amsfonts}
\usepackage{amsmath}

\begin{document}

\title{Coupled nonlinear oscillators: metamorphoses of amplitude profiles.
The case of the approximate effective equation.}
\author{Jan Kyziol$^{1)}$, Andrzej Okninski$^{2)}$ \\
Department of Mechatronics and Mechanical Engineering$^{1)}$, \\
Physics Division, Department of Management and Computer Modelling$^{2)}$, \\
Politechnika Swietokrzyska, Al. 1000-lecia PP7, \\
25-314 Kielce, Poland}
\maketitle

\begin{abstract}
We study dynamics of two coupled periodically driven oscillators. Important
example of such a system is a dynamic vibration absorber which consists of a
small mass attached to the primary vibrating system of a large mass.

Periodic solutions of the approximate effective equation are determined
within the Krylov-Bogoliubov-Mitropolsky approach to get the amplitude
profiles $A\left( \Omega \right) $.

Dependence of the amplitude $A$ of nonlinear resonances on the frequency $%
\Omega $ is much more complicated than in the case of one Duffing oscillator
and hence new nonlinear phenomena are possible. In the present paper we
study metamorphoses of the function $A\left( \Omega \right) $ induced by
changes of the control parameters.
\end{abstract}

\section{Introduction}

Coupled oscillators play important role in many scientific fields, e.g.
biology, electronics, and mechanics, see \cite%
{Awrejcewicz1991,Kozlowski1995,Kuznetsov2009} and references therein. In
this paper we analyse two coupled oscillators, one of which is driven by an
external periodic force. Important example of such system is a dynamic
vibration absorber which consists of a mass $m_{2}$, attached to the primary
vibrating system of mass $m_{1}$ \cite{DenHartog1985,Oueini1999}. Equations
describing dynamics of such system are of form:%
\begin{equation}
\left. 
\begin{array}{l}
m_{1}\ddot{x}_{1}-V_{1}\left( \dot{x}_{1}\right) -R_{1}\left( x_{1}\right)
+V_{2}\left( \dot{x}_{2}-\dot{x}_{1}\right) +R_{2}\left( x_{2}-x_{1}\right)
=f\cos \left( \omega t\right) \\ 
m_{2}\ddot{x}_{2}-V_{2}\left( \dot{x}_{2}-\dot{x}_{1}\right) -R_{2}\left(
x_{2}-x_{1}\right) =0%
\end{array}%
\right\}  \label{model1}
\end{equation}%
where $V_{1}$, $R_{1}$ and $V_{2}$, $R_{2}$ represent (nonlinear) force of
internal friction and (nonlinear) elastic restoring force for mass $m_{1}$
and mass $m_{2}$, respectively. In the present paper we do not assume that
the ratio $m_{2}/m_{1}$\ is small.

In the present paper we shall consider a simplified model:%
\begin{equation}
R_{1}\left( x_{1}\right) =-\alpha _{1}x_{1},\ V_{1}\left( \dot{x}_{1}\right)
=-\nu _{1}\dot{x}_{1}.  \label{simplified}
\end{equation}

Dynamics of coupled periodically driven oscillators is very complicated \cite%
{Awrejcewicz1991,Kozlowski1995,Kuznetsov2009}. We simplified the set
equations (\ref{model1}), (\ref{simplified}) by reducing it to the problem
of motion of two independent oscillators. More exactly, we derived the exact
fourth-order nonlinear equation for internal motion as well as approximate
second-order effective equation in \cite{Okninski2006}. Moreover, applying
the Krylov-Bogoliubov-Mitropolsky method to these equations we have computed
the corresponding nonlinear resonances (cf. \cite{Okninski2006} for the case
of the effective equation). Dependence of the amplitude $A$ of nonlinear
resonances on the frequency $\omega $ is much more complicated than in the
case of Duffing oscillator and hence new nonlinear phenomena are possible.
In the present paper we study metamorphoses of the function $A\left( \omega
\right) $ induced by changes of the control parameters.

The paper is organized as follows. In the next Section derivation of the
exact 4th-order equation for the internal motion and approximate 2nd-order
effective equations in non-dimensional form are presented. In Section 3
metamorphoses of amplitude profiles determined within the
Krylov-Bogoliubov-Mitropolsky approach for the approximate 2nd-order
effective equation are studied and the case of the standard Duffing equation
is presented as well. More exactly, theory of algebraic curves is used to
compute singular points on effective equation amplitude profiles -
metamorphoses of amplitude profiles occur in neighbourhoods of such points.
In Section 4 examples of analytical and numerical computations are presented
for the effective equation. Our results are summarized and perspectives of
further studies are described in the last Section.

\section{Exact equation for internal motion and its approximations}

In new variables, $x\equiv x_{1}$, $y\equiv x_{2}-x_{1}$, equations (\ref%
{model1}), (\ref{simplified}) can be written as:

\begin{equation}
\left. 
\begin{array}{l}
m\ddot{x}+\nu \dot{x}+\alpha x+V_{e}\left( \dot{y}\right) +R_{e}\left(
y\right) =f\cos \left( \omega t\right) \\ 
m_{e}\left( \ddot{x}+\ddot{y}\right) -V_{e}\left( \dot{y}\right)
-R_{e}\left( y\right) =0%
\end{array}%
\right\} ,  \label{model2}
\end{equation}%
where $m\equiv m_{1}$, $m_{e}\equiv m_{2}$, $\nu \equiv \nu _{1}$, $\alpha \equiv \alpha _{1}$, $V_{e}\equiv V_{2}$, 
$R_{e}\equiv R_{2}$.

Adding equations (\ref{model2}) we obtain important relation between
variables $x$ and $y$:

\begin{equation}
M\ddot{x}+\nu _{1}\dot{x}+\alpha _{1}x+m_{e}\ddot{y}=f\cos \left( \omega
t\right) ,  \label{relation}
\end{equation}%
where $M=m+m_{e}$.

We can eliminate variable $x$ in (\ref{model2}) to obtain the following
exact equation for relative motion:
\begin{equation}
\left( M\tfrac{d^{2}}{dt^{2}}+\nu \tfrac{d}{dt}+\alpha \right) \left( \mu 
\ddot{y}-V_{e}\left( \dot{y}\right) -R_{e}\left( y\right) \right) +\lambda
m_{e}\left( \nu \tfrac{d}{dt}+\alpha \right) \ddot{y}=F\cos \left( \omega
t\right) ,  \label{4th-a}
\end{equation}%
where $F=m_{e}\omega ^{2}f$,  $\mu =mm_{e}/M$  and $\lambda =m_{e}/M$ 
is a nondimensional parameter. Equations (\ref{4th-a}), (\ref%
{relation}) are equivalent to the initial equations (\ref{model1}), (\ref%
{simplified}) \cite{Okninski2006}.

In the present work we assume:

\begin{equation}
R_{e}\left( y\right) =\alpha _{e}y-\gamma _{e}y^{3},\quad V_{e}\left( \dot{y}%
\right) =-\nu _{e}\dot{y}.  \label{ReVe}
\end{equation}%
We thus get:%
\begin{equation}
\left( M\tfrac{d^{2}}{dt^{2}}+\nu \tfrac{d}{dt}+\alpha \right) \left( \mu 
\tfrac{d^{2}y}{dt^{2}}+\nu _{e}\tfrac{dy}{dt}-\alpha _{e}y+\gamma
_{e}y^{3}\right) +\lambda m_{e}\left( \nu \tfrac{d}{dt}+\alpha \right) 
\tfrac{d^{2}y}{dt^{2}}=F\cos \left( \omega t\right) .  \label{4th-b}
\end{equation}

We shall write Eq. (\ref{4th-b}) in nondimensional form. Introducing
nondimensional time $\tau $ and rescaling variable $y$:

\begin{equation}
\tau =t\bar{\omega},\ z=y\sqrt{\frac{\gamma _{e}}{\alpha _{e}}},
\label{Ndim1}
\end{equation}%
where:%
\begin{equation}
\bar{\omega}=\sqrt{\frac{\alpha _{e}}{\mu }},  \label{omega}
\end{equation}%
we get the exact equation for motion of mass $m_e$:
\begin{equation}
\left( \tfrac{d^{2}}{d\tau ^{2}}+H\tfrac{d}{d\tau }+a\right) \left( \tfrac{%
d^{2}z}{d\tau ^{2}}+h\tfrac{dz}{d\tau }-z+z^{3}\right) +\kappa \left( H%
\tfrac{d}{d\tau }+a\right) \tfrac{d^{2}z}{d\tau ^{2}}=G\tfrac{\kappa }{%
\kappa +1}\Omega ^{2}\cos \left( \Omega \tau \right) ,  \label{4th-4c}
\end{equation}%
where nondimensional constants are given by:%
\begin{equation}
h=\tfrac{\nu _{e}}{\mu \bar{\omega}},\ H=\tfrac{\nu }{M\bar{\omega}},\
\Omega =\tfrac{\omega }{\bar{\omega}},\ G=\tfrac{1}{\alpha _{e}}\sqrt{\tfrac{%
\gamma _{e}}{\alpha _{e}}}f,\ \kappa =\tfrac{m_{e}}{m},\ a=\tfrac{\alpha \mu 
}{\alpha _{e}M}.  \label{Ndim2}
\end{equation}

We shall consider hierarchy of approximate equations arising from (\ref%
{4th-4c}). For small $\kappa ,\ H,\ a$ we can reject the second term on the
left in (\ref{4th-4c}) to obtain the approximate equation:%
\begin{equation}
\left( \tfrac{d^{2}}{d\tau ^{2}}+H\tfrac{d}{d\tau }+a\right) \left( \tfrac{%
d^{2}z}{d\tau ^{2}}+h\tfrac{dz}{d\tau }-z+z^{3}\right) =\gamma \Omega
^{2}\cos \left( \Omega \tau \right) ,\qquad \left( \gamma \equiv G\tfrac{%
\kappa }{\kappa +1}\right)  \label{4th-eff}
\end{equation}%
which can be integrated partly to yield the effective equation:%
\begin{equation}
\frac{d^{2}z}{d\tau ^{2}}+h\frac{dz}{d\tau }-z+z^{3}=-\gamma \tfrac{\Omega
^{2}}{\sqrt{\left( \Omega ^{2}-a\right) ^{2}+H^{2}\Omega ^{2}}}\cos \left(
\Omega \tau +\delta \right) ,  \label{effective}
\end{equation}%
where transient states has been omitted \cite{Okninski2006}. And finally,
for $H=0,\ a=0$ we get the Duffing equation:%
\begin{equation}
\frac{d^{2}z}{d\tau ^{2}}+h\frac{dz}{d\tau }-z+z^{3}=-\gamma G\cos \left(
\Omega \tau +\delta \right) .  \label{Duffing}
\end{equation}

\section{Metamorphoses of the amplitude profiles}

We applied the Krylov-Bogoliubov-Mitropolsky (KBM) perturbation approach 
\cite{Nayfeh1981, Awrejcewicz2006} to the effective equation (\ref{effective}%
) obtaining for the $1:1$ resonance the following amplitude profile \cite%
{Okninski2006}:%
\begin{equation}
A_{eff}=\frac{\gamma \Omega ^{2}}{\sqrt{\left( h^{2}\Omega ^{2}+\left(
1+\Omega ^{2}-\frac{3}{4}A_{eff}^{2}\right) ^{2}\right) \left( \left( \Omega
^{2}-a\right) ^{2}+H^{2}\Omega ^{2}\right) }}.  \label{Aeff}
\end{equation}%
Now, for $H=0,\ a=0$, we obtain the amplitude profile for the Duffing
equation (\ref{Duffing}):%
\begin{equation}
A_{D}=\frac{\gamma }{\sqrt{\left( h^{2}\Omega ^{2}+\left( 1+\Omega ^{2}-%
\frac{3}{4}A_{D}^{2}\right) ^{2}\right) }}.  \label{ADuff}
\end{equation}

It is well known that dependence of the function $A_{D}$, cf. (\ref{ADuff}),
on control parameters $\gamma $, $h$ is rather simple. On the other hand,
dependence of the amplitude profile $A_{eff}\left( \Omega \right) $ on
control parameters $\gamma $, $h$, $a$, $H$ is more complicated and thus $%
A_{eff}\left( \Omega \right) $\ can describe new nonlinear phenomena. In the
next Section we shall study possible metamorphoses of $A_{D}$, $A_{eff}$
induced by changes of control parameters, the more complicated case of the
4th-order exact equation (\ref{4th-4c}) will be treated elsewhere.

Equations (\ref{ADuff}), (\ref{Aeff}) define the corresponding amplitude
profiles implicitly. Such amplitude profiles can be classified as planar
algebraic curves. Firstly, we shall collect useful theorems on implicit
functions which will be used below.

Let us write equations (\ref{Aeff}), (\ref{ADuff}) as $L_{e}\left(
Y_{e},X\right) =0$ and $L_{D}\left( Y_{D},X\right) =0$, respectively, where $%
X\equiv \Omega ^{2}$, $Y\equiv A^{2}$:%
\begin{eqnarray}
\left( h^{2}X+\left( 1+X-\tfrac{3}{4}Y_{e}\right) ^{2}\right) \left( \left(
X-a\right) ^{2}+H^{2}X\right) Y_{e}-\gamma ^{2}X^{2} &=&0,  \label{Feff} \\
\left( h^{2}X+\left( 1+X-\tfrac{3}{4}Y_{D}\right) ^{2}\right) Y_{D}-\gamma
^{2} &=&0.  \label{FD}
\end{eqnarray}

It follows from general theory of implicit functions \cite{Spivak1965,Gibson1998} that
conditions for critical points of $Y\left( X\right) $ read:
\begin{equation}
L\left( Y,X\right) =0,\qquad \frac{\partial L\left( Y,X\right) }{\partial X}%
=0\qquad \left( \frac{\partial L\left( Y,X\right) }{\partial Y}\neq 0\right)
.  \label{Cr1}
\end{equation}%
Moreover, critical points of the inverse function $X\left( Y\right) $ are
given by:%
\begin{equation}
L\left( Y,X\right) =0,\qquad \frac{\partial L\left( Y,X\right) }{\partial Y}%
=0\qquad \left( \frac{\partial L\left( Y,X\right) }{\partial X}\neq 0\right)
.  \label{Cr2}
\end{equation}

It may happen that in some points $\left( X_{0},Y_{0}\right) $ we have:%
\begin{equation}
L\left( Y,X\right) =0,\qquad \frac{\partial L\left( Y,X\right) }{\partial X}%
=0,\qquad L\left( Y,X\right) =0.  \label{Singular}
\end{equation}

Such points are referred to as singular points of algebraic curve $L\left(
Y,X\right) =0$ because they are in some sense exceptional.

\subsection{The case of the Duffing equation}

Singular points of the algebraic curve defined by (\ref{FD}) are given by:%
\begin{eqnarray}
\frac{\partial L_{D}}{\partial X} &=&0,  \label{SD1} \\
\frac{\partial L_{D}}{\partial Y} &=&0.  \label{SD2}
\end{eqnarray}

The set of equations (\ref{FD}), (\ref{SD1}), (\ref{SD2}) can be written as:%
\begin{eqnarray}
h^{2}XY+Y\left( 1+X-\tfrac{3}{4}Y\right) ^{2}-\gamma ^{2} &=&0,  \label{D1}
\\
h^{2}Y+2Y+2YX-\tfrac{3}{2}Y^{2} &=&0,  \label{D2} \\
h^{2}X+1+2X-3Y+X^{2}-3YX+\tfrac{27}{16}Y^{2} &=&0,  \label{D3}
\end{eqnarray}%
where $X,Y$ are positive.

General solution reads:%
\begin{equation}
\left\{ 
\begin{array}{l}
X=-\tfrac{1}{2}h^{2}-1+\tfrac{3}{4}Y\qquad \left( Y\neq 0\right) \\ 
Y=\tfrac{1}{6}h^{2}+\tfrac{2}{3}\qquad \left( h\neq 0\right) \\ 
\gamma ^{2}=-\tfrac{1}{48}h^{6}-\tfrac{1}{6}h^{4}-\tfrac{1}{3}h^{2}%
\end{array}%
\right. .  \label{Dsol}
\end{equation}

It follows that $Y>0$, $X<0$ and $\gamma ^{2}\leq 0$ and thus the system of
equations (\ref{D1}), (\ref{D2}), (\ref{D3}) has no acceptable solutions
since we assume that $h,\ \gamma $ are real and $X,\ Y$ are non-negative.

\subsection{The case of the effective equation}

Singular points of the algebraic curve defined by (\ref{Feff}) are given by
equations:%
\begin{eqnarray}
\frac{\partial L_{e}}{\partial X} &=&0,  \label{Se1} \\
\frac{\partial L_{e}}{\partial Y} &=&0.  \label{Se2}
\end{eqnarray}

It follows from (\ref{Feff}) and (\ref{Se2}) that either of equations must
hold: 
\begin{subequations}
\label{CASES}
\begin{eqnarray}
16X^{2}+16h^{2}X+32X+16-48YX-48Y+27Y^{2} &=&0,  \label{case1} \\
\left( X-a\right) ^{2}+H^{2}X &=&0,  \label{case2}
\end{eqnarray}%
where $X,Y$ are positive.

Let us start with Eq.(\ref{case2}). In this case we obtain from (\ref{Feff}%
), (\ref{Se1}) and (\ref{case2}) the following rather special solution: 
\end{subequations}
\begin{equation}
X=a,H=0,\gamma =0\qquad \left( Y,h,a\text{ - arbitrary}\right) .
\label{solution1}
\end{equation}

Let us now consider more general Eq.(\ref{case1}). We can treat $X$ as
arbitrary. Then we obtain two solutions for $Y$:%
\begin{eqnarray}
Y &=&\tfrac{8}{9}+\tfrac{8}{9}X-\tfrac{4}{9}\sqrt{1+2X+X^{2}-3h^{2}X},
\label{Y1} \\
Y &=&\tfrac{8}{9}+\tfrac{8}{9}X+\tfrac{4}{9}\sqrt{1+2X+X^{2}-3h^{2}X},
\label{Y2}
\end{eqnarray}%
where the inequality%
\begin{equation}
1+2X+X^{2}-3h^{2}X\geq 0,  \label{inequality}
\end{equation}%
must hold. This means that for a chosen value of $X$ the parameter $h$ must
obey%
\begin{equation}
h^{2}\leq \tfrac{\left( X+1\right) ^{2}}{3X}.  \label{ineqh}
\end{equation}

Solving equations (\ref{Feff}), (\ref{Se1}), and (\ref{Y1}) or (\ref{Y2}) we
get:%
\begin{eqnarray}
Y &=&\tfrac{8}{9}+\tfrac{8}{9}X\pm \tfrac{4}{9}U\left( X\right) ,  \label{Y}
\\
a &=&Z_{1}\tfrac{X}{\left( h^{2}X+X^{2}+2X+1\right) ^{\frac{3}{2}}h},
\label{a} \\
H &=&Z_{2}\tfrac{1}{\left( h^{2}X+X^{2}+2X+1\right) ^{\frac{3}{2}}h},
\label{H}
\end{eqnarray}%
where $U$ and $Z_{1}$, $Z_{2}$ are given by:%
\begin{eqnarray}
U\left( X\right) &=&\sqrt{1+2X+X^{2}-3h^{2}X}  \label{U} \\
Z_{1} &=&\sqrt{w_{1}\left( X\right) \pm w_{2}\left( X\right) U\left(
X\right) },  \label{Z1} \\
Z_{2} &=&\sqrt{w_{3}\left( X\right) \pm w_{4}\left( X\right) U\left(
X\right) },  \label{Z2}
\end{eqnarray}%
with%
\begin{equation}
w_{1}\left( X\right)
=a_{6}X^{6}+a_{5}X^{5}+a_{4}X^{4}+a_{3}X^{3}+a_{2}X^{2}+a_{1}X+a_{0},
\label{w1}
\end{equation}%
\begin{equation}
\left\{ 
\begin{array}{l}
a_{6}=16h^{2}, \\ 
a_{5}=48h^{4}+96h^{2}, \\ 
a_{4}=48h^{6}+192h^{4}+240h^{2}+6\gamma ^{2}, \\ 
a_{3}=16h^{8}+96h^{6}+288h^{4}+\left( 320+102\gamma ^{2}\right)
h^{2}+24\gamma ^{2}, \\ 
a_{2}=48h^{6}+192h^{4}+\left( 162\gamma ^{2}+240\right) h^{2}+36\gamma ^{2},
\\ 
a_{1}=\left( 54\gamma ^{2}+48\right) h^{4}+\left( 18\gamma ^{2}+96\right)
h^{2}+24\gamma ^{2}, \\ 
a_{0}=\left( 16-42\gamma ^{2}\right) h^{2}+6\gamma ^{2},%
\end{array}%
\right.  \label{coeffw1}
\end{equation}%
\begin{equation}
w_{2}\left( X\right) =b_{3}X^{3}+b_{2}X^{2}+b_{1}X+b_{0},  \label{w2}
\end{equation}%
\begin{equation}
\left\{ 
\begin{array}{l}
b_{3}=6\gamma ^{2}, \\ 
b_{2}=-51\gamma ^{2}h^{2}+18\gamma ^{2}, \\ 
b_{1}=-9\gamma ^{2}h^{4}-12\gamma ^{2}h^{2}+18\gamma ^{2}, \\ 
b_{0}=39\gamma ^{2}h^{2}+6\gamma ^{2},%
\end{array}%
\right.  \label{coeffw2}
\end{equation}
\begin{equation}
w_{3}\left( X\right)
=c_{7}X^{7}+c_{6}X^{6}+c_{5}X^{5}+c_{4}X^{4}+c_{3}X^{3}+c_{2}X^{2}+c_{1}X+c_{0},
\label{w3}
\end{equation}%
\begin{equation}
\left\{ 
\begin{array}{l}
c_{7}=-32h^{2}, \\ 
c_{6}=-96h^{4}-192h^{2}, \\ 
c_{5}=-96h^{6}-384h^{4}-480h^{2}+32Z_{1}h, \\ 
c_{4}=-32h^{8}-192h^{6}-576h^{4}+64Z_{1}h^{3}+ \\ 
\qquad \left( -640-42\gamma ^{2}\right) h^{2}+128Z_{1}h+6\gamma ^{2}, \\ 
c_{3}=-96h^{6}+32Z_{1}h^{5}+\left( -384+54\gamma ^{2}\right) h^{4}+ \\ 
\qquad 128Z_{1}h^{3}+\left( -480+18\gamma ^{2}\right)
h^{2}+192Z_{1}h+24\gamma ^{2}, \\ 
c_{2}=162\gamma ^{2}h^{2}+64Z_{1}h^{3}+128Z_{1}h-192h^{2}+36\gamma
^{2}-96h^{4}, \\ 
c_{1}=\left( 102\gamma ^{2}-32\right) h^{2}+32Z_{1}h+24\gamma ^{2}, \\ 
c_{0}=6\gamma ^{2},%
\end{array}%
\right.  \label{coeffw3}
\end{equation}%
\begin{equation}
w_{4}\left( X\right) =b_{0}X^{3}+b_{1}X^{2}+b_{2}X+b_{3}.  \label{w4}
\end{equation}

\section{Analytical and numerical computations}

It follows from solutions obtained in the preceding Section that we can
control position of a singular point. More exactly, we choose a value of $X$
and then $h$ fulfilling inequality (\ref{ineqh}) can be chosen as well. Next
we specify $\gamma $ and then $Y$, $a$, $H$ are computed from Eqs. (\ref{Y}%
), (\ref{a}), (\ref{H}). In this process the position of the singular point $%
\left( X,\ Y\right) $ and values of control parameters $H$, $h$, $\gamma $, $%
a$ are determined (provided that the solutions are real) .

\begin{figure}[h!]
\begin{equation*}
\includegraphics[width= 8 cm]{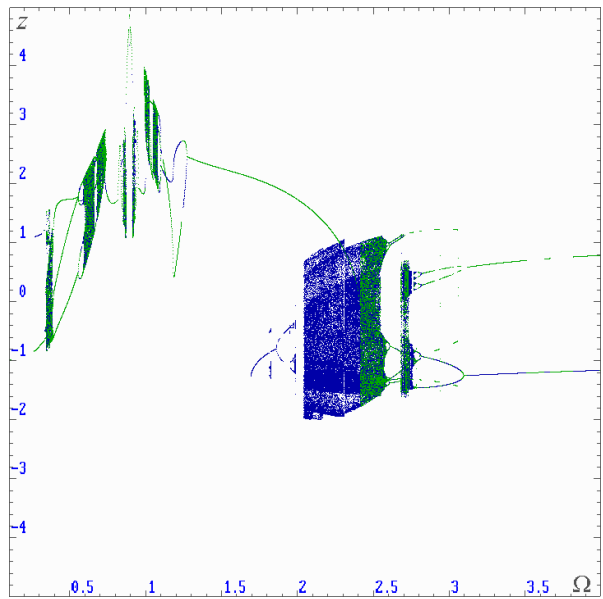}
\end{equation*}%
\caption{Bifurcation diagram for the effective equation, $h=0.4$, $\gamma=2.5$, $a=0.8$, $H=0.04$.}
\label{F1}
\end{figure}

\begin{figure}[h!]
\begin{equation*}
\includegraphics[width= 8 cm]{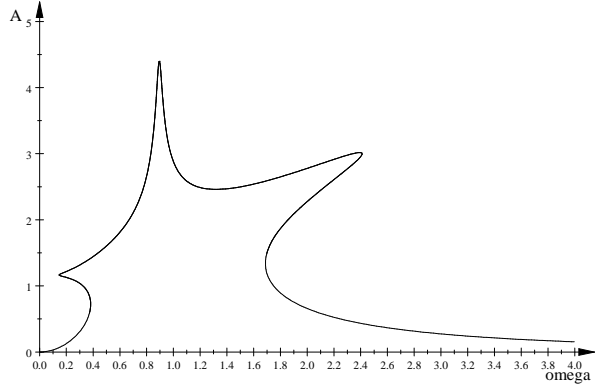}
\end{equation*}%
\caption{Amplitude profile $A(\Omega)$, $h=0.4$, $\gamma=2.5$, $a=0.8$, $H=0.04$.}
\label{F2}
\end{figure}

Bifurcation diagram for the effective equation (\ref{effective}) for the
following values of control parameters $H=0.04$, $h=0.4$, $a=0.8$, $\gamma
=2.5$ is shown in Fig. \ref{F1} (cf. Fig. 1\textsl{\ }in \cite{Okninski2006}) 
where colours mark different initial conditions.
Position of the $1:1$ resonance agrees well with the amplitude profile, 
computed for the same parameters, cf. Fig. \ref{F2} and discussion in \cite{Okninski2006}.

\begin{figure}[h!]
\begin{equation*}
\includegraphics[width= 8 cm]{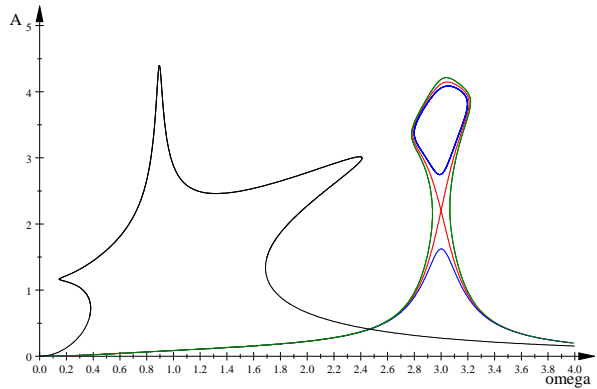}
\end{equation*}%
\caption{$A(\Omega)$ in the singular point and in its neighbourhood, $h=0.8$, $\gamma=1.5$, $a=9.0720$, and $H=0.27$ (green), $H=0.2995$ (red), $H=0.33$ (blue).}
\label{F3}
\end{figure}

We shall now compute coordinates of a singular point using equations (\ref{Y}%
), (\ref{a}), (\ref{H}). At first we choose the value of $X$ as $X=9$ $%
\left( \Omega =3\right) $. Then we can select any value of $h$ obeying
inequality (\ref{ineqh}). We thus put $h=0.8$ to get from Eq. (\ref{Y1}) $%
Y=4.\,8466$ $\left( A=2.2015\right) $. Next we choose $\gamma =1.5$ to
compute from (\ref{a}), (\ref{H}) $a=9.0720$, $H=0.2995$. In Fig. \ref{F3} below we show
amplitude profiles computed from Eq. (\ref{Aeff}) for critical parameter
values $\left( H,\ h,\ a,\ \gamma \right) =\left( 0.2995,\ 0.8,\ 9.0720,\
1.5\right) $, and for two more values of $H$, $H>H_{cr}$ and $H<H_{cr}$.
\bigskip 

\begin{figure}[h!]
\begin{equation*}
\includegraphics[width= 7.5 cm]{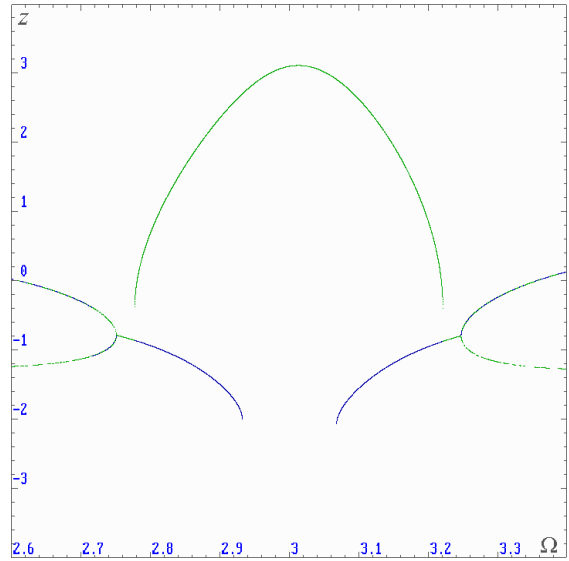}
\end{equation*}
\caption{Bifurcation diagram, $h=0.8$, $\gamma=1.5$, $a=9.0720$, $H=0.27$.}
\label{F4}
\end{figure}

\begin{figure}[h!]
\begin{equation*}
\includegraphics[width= 7.5 cm]{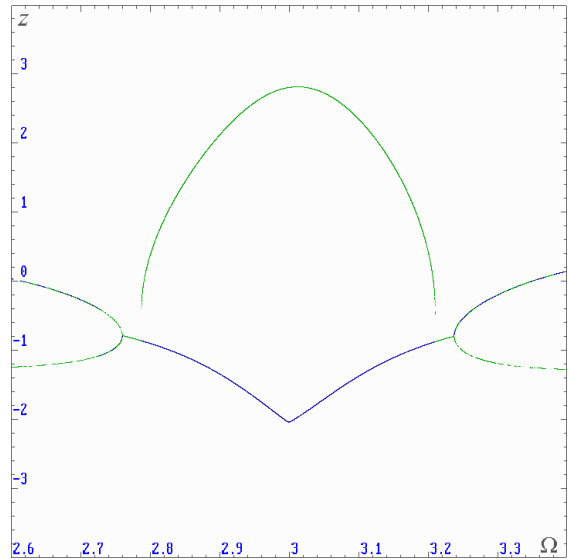}
\end{equation*}%
\caption{Critical diagram, $h=0.8$, $\gamma=1.5$, $a=9.0720$, $H=0.3019$.}
\label{F5}
\end{figure}

The critical (red) curve intersects itself in singular point $\left( X,\ Y \right)
=\left( 4.\,8466,\ 9\right) $ or $\left( A,\ \Omega \right) =\left( 2.\,2015,\ 3\right) $. 
Green curve corresponds to $H=0.27$ while
blue curve has been computed for $H=0.33$ (other parameter values
unchanged). The initial amplitude from Fig. \ref{F1} was also shown (black
curve). The first bifurcation diagram, cf. Fig. \ref{F4}, was computed for $%
H=0.27$ and corresponds to the green curve in Fig. \ref{F3}. We note that
the small branch of the $1:1$ resonance is discontinuous in agreement with
the amplitude profile shown in Fig. \ref{F3} (green curve). The next
bifurcation digram, Fig. \ref{F5}, has been computed for critical value $H=0.3019$ determined numerically 
from Eq. (\ref{effective}) (this differs slightly from the critical value $H=0.2995$ determined from the 
KBM solution as described above).

And finally, the last bifurcation diagram was computed for $H=0.33$ - and
again the small branch of the $1:1$ resonance is continuous.

\begin{figure}[h!]
\begin{equation*}
\includegraphics[width= 7.5 cm]{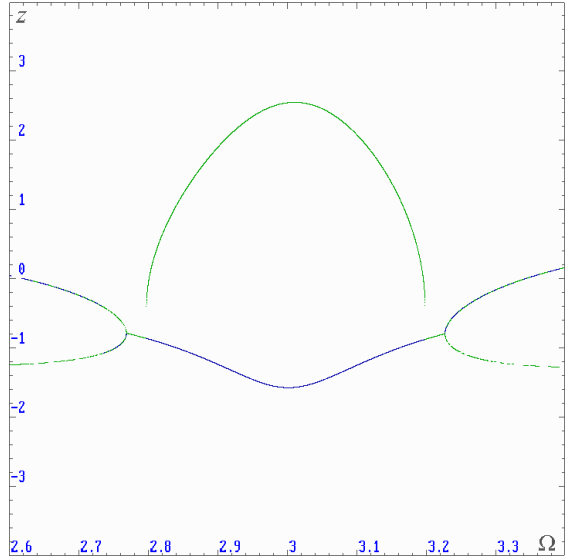}
\end{equation*}%
\caption{Bifurcation diagram, $h=0.8$, $\gamma=1.5$, $a=9.0720$, $H=0.33$.}
\label{F6}
\end{figure}

It follows from results presented in Section that for $X=9$, $h=0.8$, $%
\gamma =1.5$ there is another singular point. Indeed, we can compute $Y$
from another of equations (\ref{Y}) to get from Eqs. (\ref{a}), (\ref{H}) $%
Y=12.9311$ $\left( A=3.5960\right) $, $a=9.1213$, $H=0.5158$.

\begin{figure}[h!]
\begin{equation*}
\includegraphics[width= 8 cm]{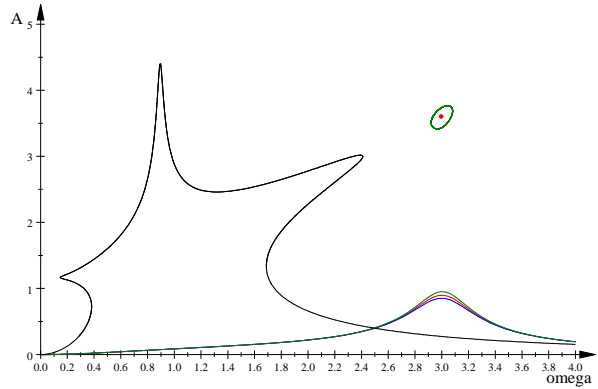}
\end{equation*}%
\caption{$A(\Omega)$ in the singular point and in its neighbourhood, $h=0.8$, $\gamma=1.5$, $a=9.1213$, and $H=0.49$ (green), $H=0.5158$ (red), $H=0.54$ (blue).}
\label{F7}
\end{figure}

In Fig. \ref{F7} amplitude profiles computed from Eq. (\ref{Aeff}) for
critical parameter values $\left( H,\ h,\ a,\ \gamma \right) =\left(
0.5158,\ 0.8,\ 9.1213,\ 1.5\right) $ and for two other values of $H$, $%
H<H_{cr}$ and $H>H_{cr}$ have been shown. Bifurcation diagrams for $H = 0.49$
and $H = 0.54$ are shown below.
\begin{figure}[h!]
\begin{equation*}
\includegraphics[width= 7.5 cm]{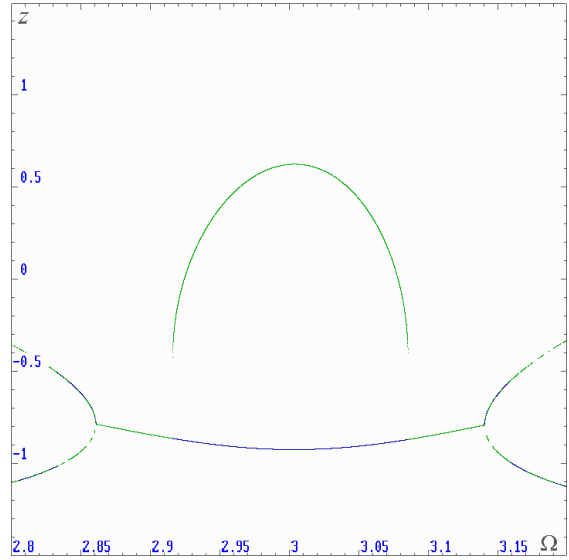}
\end{equation*}%
\caption{Bifurcation diagram, $h=0.8$, $\gamma=1.5$, $a=9.1213$, $H=0.49$.}
\label{F8}
\end{figure}
\begin{figure}[h!]
\begin{equation*}
\includegraphics[width= 7.5 cm]{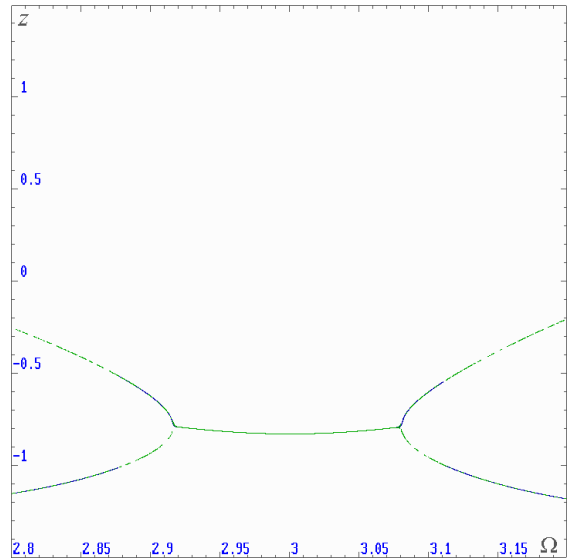}
\end{equation*}%
\caption{Bifurcation diagram, $h=0.8$, $\gamma=1.5$, $a=9.0720$, $H=0.54$.}
\label{F9}
\end{figure}

\section{Summary and discussion}

In this work we have studied metamorphoses of amplitude profiles for the
effective equation, describing approximately dynamics of two coupled
periodically driven oscillators. Our analysis has been analytical although
based on the approximate KBM method. 

Theory of algebraic curves has been used to compute singular points on
effective equation amplitude profiles. It follows from general theory that
metamorphoses of amplitude profiles occur in neighbourhoods of such points.
In Section 3 we have computed analytically positions of singular points for
the amplitude profiles $A\left( \Omega \right) $\ determined within the
Krylov-Bogoliubov-Mitropolsky approach for the approximate 2nd-order
effective equation (\ref{effective}). In the first case the singular point
corresponds to self-intersection of $A\left( \Omega \right) $, see Fig. \ref%
{F3}, while in the second case it is a isolated point, cf. Fig. \ref{F7}. 

It is interesting that the solution described in Section 3 permits control of
position of singular point: we choose arbitrary value of variable $X$ $%
\left( X^{2}=\Omega \right) $, then value of the parameter $h$ obeying
inequality (\ref{inequality}) is selected. Finally the value of the control
parameter $\gamma $ is chosen and $Y$, $a$, $H$ are computed from Eqs. (\ref%
{Y}), (\ref{a}), (\ref{H}); it should be stressed that we have not come
across any difficulties to obtain real solutions. We hope to carry full
analysis of conditions guaranteeing existence of real solutions in our
future papers. As a by-product we have demonstrated that there are no
singular points for $A\left( \Omega \right) $ computed for the Duffing
equation in agreement with well established numerical experience. 

We have also computed numerically bifurcation diagrams in the neighbourhoods of
singular points and indeed dynamics of the effective equation (\ref
{effective}) changes according to metamorphoses of the
corresponding amplitude profiles. 

In our future work we are going to study singular points of the amplitude
profiles computed for the exact equation (\ref{4th-4c}).

\end{document}